\documentclass{llncs}
\usepackage[T1]{fontenc}

\usepackage{amssymb}
\usepackage{graphicx}
\usepackage{enumerate}
\usepackage{hyperref}
\usepackage{url}

\usepackage[ruled,vlined]{algorithm2e}
\usepackage{amsmath}
\usepackage{amsfonts}
\usepackage{alltt}

\usepackage{mathptmx}
\usepackage{stmaryrd}

\usepackage{pgfplots}
\usepackage{tikz}

\usepackage{url}

\urldef{\mailsa}\path|dignatov@hse.ru|
\newcommand{\keywords}[1]{\par\addvspace\baselineskip
\noindent\keywordname\enspace\ignorespaces#1}

\newcommand{\BV}{\underline{{\mathfrak B}}}

\begin{document}

\mainmatter  

\title{RAPS: A Recommender Algorithm Based on Pattern Structures}

\titlerunning{RAPS: A Recommender Algorithm based on Pattern Structures}

%
%
\author{Dmitry I. Ignatov\inst{1} 
 \and Denis Kornilov\inst{1}}
\authorrunning{}

\institute{National Research University Higher School of Economics\\
\mailsa\\
\url{http://www.hse.ru}
}

%
%

\toctitle{RAPS: Recommender Algorithm based on Pattern Structures}
\tocauthor{Ignatov and Kornilov}
\maketitle

\begin{abstract}
We propose a new algorithm for recommender systems with numeric ratings which is based on Pattern Structures (RAPS). As the input the algorithm takes rating matrix, e.g., such that it contains movies rated by users. For a target user, the algorithm returns a rated list of items (movies) based on its previous ratings and ratings of other users. We compare the results of the proposed algorithm in terms of precision and recall measures with Slope One, one of the state-of-the-art item-based algorithms, on Movie Lens dataset and RAPS demonstrates the best or comparable quality.

\end{abstract}

\section{Introduction and related work}

Formal Concept Analysis (FCA)\cite{Ganter:1999:FCA} is a powerful algebraic framework for knowledge representation and processing \cite{Poelmans:2013a,Poelmans:2013b}. However, in its original formulation it deals with mainly Boolean data. Even though original numeric data can be represented by so called multi-valued context, it requires concept scaling to be transformed to a plain context (i.e. a binary object-attribute table). There are several extensions of FCA to numeric setting like Fuzzy Formal Concept Analysis \cite{Belohlavek:2011,Poelmans:2014}. In this paper, to recommend particular user items of interest we use Pattern Structures, an extension of FCA to deal with data that have ordered descriptions. In fact, we use interval pattern structures that were proposed in \cite{Ganter:2001} and successfully applied, e.g., in gene expression data analysis \cite{Kaytoue:2011}.

The task of recommending items to users according to their preferences expressed by ratings of previously used items became extremely popular during the last decade partially because of famous NetFlix 1M\$ competition \cite{Bell:2007}. Numerous algorithms were proposed to this end. In this paper we will mainly study item-based approaches. Our main goal is to see whether FCA-based approaches are directly applicable to the setting of recommender systems with numeric data. Previous approaches used concept lattices for navigation through the recommender space and allowed to recommend relevant items faster than online computation in user-based approach, however it requires expensive offline computations and a substantial storage space \cite{duBoucher:2006}.  Another approach tries to effectively use Boolean factorisation based on formal concepts and follows user-based k-nearest neighbours strategy \cite{Ignatov:2014}. A parameter-free approach that exploits a neighbourhood of the object concept for a particular user also proved its effectiveness \cite{Alqadah:2014} but it has a predecessor based on object-attribute biclusters \cite{Ignatov:2012b} that also capture the neighbourhood of every user and item pair in an input formal context. However, it seems that within FCA framework item-based techniques for data with ratings have not been proposed so far. So, the paper bridges the gap.

The paper is organised as follows. In Section \ref{sec:defs}, basic FCA definitions and interval pattern structures are introduced. Section \ref{sec:algs} describes SlopeOne \cite{Lemire:2005} and RAPS with examples. In Section \ref{sec:exp}, we provide the results of experiments with time performance and precision-recall evaluation for MovieLens dataset. Section \ref{sec:end} concludes the paper.

\section{Basic definitions} \label{sec:defs}

\paragraph{Formal Concept Analysis.}

First, we  recall several basic notions of Formal Concept Analysis
(FCA) \cite{Ganter:1999:FCA}. Let $G$ and $M$ be sets, called the
set  of objects and attributes, respectively, and let $I$ be a
relation $I\subseteq G\times M$: for $g\in G, \ m\in M$,  $gIm$
holds  iff the  object  $g$  has  the attribute $m$. The triple
$\mathbb{K}=(G,M,I)$ is called a {\it (formal) context}. If $A\subseteq G$,
$B\subseteq  M$ are arbitrary subsets, then the {\it Galois
connection} is given by the following {\it derivation operators}:

\begin{eqnarray}
\begin{array}{c}
A' = \{m\in M\mid gIm \ {\rm for\ all}\ g\in A\}, \\
B' = \{g\in G\mid gIm \ {\rm for\ all}\ m\in B\}.
\end{array}
\end{eqnarray}

The pair $(A,B)$, where $A\subseteq G$, $B\subseteq M$, $A' = B$,
and $B' = A$ is called a {\it (formal) concept (of the context
$K$)} with {\it extent} $A$ and {\it intent} $B$ (in this case we
have also $A'' = A$ and $B'' = B$).

The concepts, ordered by $(A_1,B_1)\geq (A_2,B_2) \iff A_1\supseteq A_2$ form a complete  lattice, called \emph{the concept lattice}  $\BV(G,M,I)$.

\paragraph{Pattern Structures.}

 Let  $G$ be a set of objects and $D$ be a set of all possible object descriptions. Let $\sqcap$ be a similarity operator. It helps to work with objects that have non-binary attributes like in traditional FCA setting, but those that have complex descriptions like intervals or graphs. Then $(D,\sqcap)$  is a meet-semi-lattice of object descriptions. Mapping  $\delta: G \to D$ assigns an object $g$ the description $d \in (D, \sqcap)$.    

A triple  $(G,(D,\sqcap),\delta)$ is a pattern structure. Two operators $( \cdot)^\square$ define Galois connection between $(2^G, \subseteq)$ and $(D,\sqcap)$:

\begin{eqnarray}
A^\square =\bigsqcap\limits_{g \in A} \delta(g)    \mbox{ for } A \subseteq G	\label{op1}\\
d^\square =\{g \in G | d \sqsubseteq \delta(g) \}   \mbox{ for }  d \in (D,\sqcap), \mbox{ where }\label{op2} \\
d \sqsubseteq \delta(g) \iff d \sqcap \delta(g)=d	.\nonumber
\end{eqnarray}

For a set of objects $A$ operator~\ref{op1} returns the common description (pattern) of all objects from $A$. For a description $d$ operator~\ref{op2} returns the set of all objects that contain $d$.

A pair $(A,d)$ such that $A\subseteq G$ and $d \in (D,\sqcap)$ is called a pattern concept of the pattern structure $(G,(D,\sqcap),\delta)$ iff $A^\square=d$ and $d^\square=A$. In this case $A$ is called a pattern extent and $d$ is called a pattern intent of a pattern concept  $(A, d)$. 
Pattern concepts are partially ordered by $(A_1,d_1) \leq (A_2,d_2 ) \iff A_1 \subseteq A_2 (\iff d_2 \sqsubseteq d_1)$. The set of all pattern concepts forms a complete lattice called a pattern concept lattice.

\paragraph{Intervals as patterns.}

It is obvious that similarity operator on intervals should fulfill the following condition:  two intervals should belong to an interval that contains them. Let this new interval be minimal one that contains two original intervals. Let  $[a_1,b_1]$ and $[a_2,b_2]$ be two intervals such that $a_1,b_1, a_2, b_2 \in \mathbb R$, $a_1\leq b_1$ and $a_2 \leq b_2$, then their similarity is defined as follows:
 
$$[a_1,b_1 ] \sqcap [a_2,b_2 ]=[\min(a_1,a_2 ), \max(b_1,b_2 )].$$

Therefore 
\begin{eqnarray*}
[a_1,b_1 ] \sqsubseteq [a_2,b_2 ] \iff [a_1,b_1 ] \sqcap [a_2,b_2 ]=[a_1,b_1 ]  \\
\iff \big[ \min (a_1, a_2) , \max (b_1, b_2) \big]  =[ a_1, b_1]\\
\iff a_1 \leq  a_2  \mbox{ and }  b_1 \geq b_2 \iff [a_1,b_1 ] \supseteq [a_2,b_2] \\
\end{eqnarray*}

Note that $a \in \mathbb R$ can be represented by $[a,a]$.

\paragraph{Interval vectors as patterns.}

Let us call p-adic vectors of intervals as interval vectors. In this case for two interval vectors of the same dimension $e=\langle [a_i,b_i] \rangle_{i \in [1,p]}$ and $f= \langle [c_i,d_i] \rangle _{i\in [1,p]}$ we define similarity operation via the intersection of the corresponding components of interval vectors, i.e.:

$$e \sqcap f=\langle [a_i,b_i ]\rangle_{i \in [1,p]} \sqcap \langle [c_i,d_i ] \rangle_{i\in [1,p]} \iff e \sqcap f=\langle [a_i,b_i] \sqcap [c_i,d_i] \rangle_{i \in [1,p] }$$

Note that interval vectors are also partially ordered:
$$e \sqsubseteq f \iff\langle [a_i,b_i ] \rangle _{i\in [1,p] } \sqsubseteq \langle [c_i,d_i ] \rangle_{i \in [1,p]} \iff  [a_i,b_i ] \sqsubseteq [c_i,d_i ]$$  for all  $i \in [1,p]$.

\section{Recommender Algorithms} \label{sec:algs}

\subsection{Slope One}

Slope One is one of the common approaches to recommedations  based on collaborative filtering. However, it  demonstrates comparable quality with more complex and resource demanding algorithms \cite{Lemire:2005}. As it was shown in \cite{Cacheda:2011}, SlopeOne has the highest recall on MovieLens and Netflix datasets and acceptable level of precision: ``Overall, the algorithms that present the best results with these metrics are SVD techniques, tendencies-based and slope one (although its precision is not outstanding).''

We use this algorithm for comparison purposes. 

Slope One deals with rating matrices as input data. In what follows the data contains   movies ratings by different users. That is $M=\{m_1,m_2,\ldots,m_n\}$ is a set of movies, $U=\{u_1,u_2,\ldots,u_k\}$ is a set of users. The rating matrix can be represented by many-valued formal context $(U,M,R, I)$, where $R=\{1,2,3,4,5, * \}$ is a set of possible ratings and a triple $(u,m,r) \in I$ means that the user $u$ marked by the rating $r$ the movie $m$. Whenever it is suitable we also use $r_{ij}$ notation for rating of movie $m_j$ by user $u_i$.

 In case a user $u$ has not rated a movie $m$,  we use $m(u)=r=*$, i.e. missing rating.

Let us describe the algorithm step by step. 

\begin{enumerate}
\item  The algorithm takes a many valued context of all users' ratings, the target user  $u_t$ for which it generates recommendations. It also requires $left\_border$ and $right\_border$ for acceptable level of ratings, i.e. if one wants to receive all movies with ratings between 4  and 5,  then left and right borders should be 4 and 5 respectively. The last pair of parameters: one needs to set up minimal and maximal scores ($min\_border$ and $max\_border$) that are acceptable for our data. It means that if the algorithm predicts rating 6.54 as a score and maximal score is  bounded by 5, then 6.54 should be treated as 5.

\item The algorithm finds the set of all movies evaluated by the target user $S(u_t)$. 
\item For every non-evaluated movie  $m_j \in  M \setminus S(u_t)$  by $u_t$ execute step 4), and by so doing calculate the predicted rating  for the movie $m_j$. After that go to step 5).
\item For every evaluated movie  $m_i \in S(u_t)$ by  $u_t$ calculate $S_{j,i} (U\setminus \{u_t\})$, the set of users  that watched and evaluated movies $m_i$ and $m_j$. In case $S_{j,i} (U\setminus \{u_t\})$ is non-empty, that is $|S_{j,i} (U\setminus \{u_t \})|>0$, calculate the deviation: $dev_{j,i}=\sum\limits_{u_k\in S_{j,i} (U\setminus \{u_t\})}\frac{r_{k,j}-r_{k,i}}{|S_{j,i} (U\setminus \{u_t \})|}$ and add $i$ to $R_j$. 

After all current deviations found, calculate the predicted rating: $P(u_t )_j=\frac{1}{|R_j |} \sum\limits_{i\in R_j}(dev_{j,i}+r_{t,i})$, where  $R_j=\{ i  |  m_i\in S(u_t ),i\neq j, |S_{j,i} (U\setminus \{u_t \})|>0 \}$. In case  $R_j$ is empty, the algorithm cannot make a prediction.
\item By this step Slope One found all predicted ratings $P(u_t)$ for movies from $M\setminus S(u_t)$. 

The algorithm recommends all movies with predicted ratings in the preferred range  $left\_border \leq P(u_t )_j\leq right\_border$, taking into account minimal and maximal allowed values. 
\end{enumerate}

If one needs top-$N$ ranked items, she can sort the predicted scores from the resulting set in decreasing order and select first $N$ corresponding movies.

\noindent\textbf{Example 1.}

Consider execution of Slope One on the dataset from Table~\ref{tab:SOex}.

\begin{table}
\caption{Example of data for Slope One}\label{tab:SOex}
	\centering
		\begin{tabular}{|c|c|c|c|}
		\hline
	user$\backslash$movie &	$m_1$	 & $m_2 $& $	m_3$\\
			\hline
$u_1$	& 5	& 3 &	2\\
$u_2$	& 3 &	4 &	*\\
$u_3$ &	* &	2	& 5\\
			\hline
		\end{tabular}

\end{table}

Let us try to predict the rating for $u_3$ and movie $m_1$.

\begin{enumerate}

\item Let  $left\_border=4$, $right\_border=5$, $min\_border=1$, and $max\_border=5$.
\item We find $S(u_3 )=\{m_2  ,m_3\}$, the set of evaluated movies by the target user. 
\item $M\setminus S(u_3 )=\{m_1\}$

\item $S_{1,2} (U\setminus \{u_3 \})=\{u_1,u_2\}$

$dev_{1,2}=\frac{(r_{1,1}-r_{1,2})+(r_{2,1}-r_{2,2})}{(|\{u_1,u_2\}|)}=((5-3)+(3-4))/2=0.5$

$S_{1,3} (U\setminus \{u_3 \})=\{u_1\}$

$dev_{1,3}=(r_{1,1}-r_{1,3})/(|\{u_1\}|)=(5-2)/1=3$

$R_1=\{2,3\}$

$P(u_3)_1=1/|R_j |  (dev_{1,2}+r_{3,2}+dev_{1,3}+r_{3,3} )=1/2 (0.5+2+3+5)=5.25$

\item Taking into account the maximal rating boundary, the algorithm predicts 5 for movie $m_1$, and therefore recommends user $u_3$ to watch it.

\end{enumerate}

\subsection{RAPS}

Our approach, RAPS (Recommender Algorithm based on Pattern Structures), works with the same many valued context as Slope One.

Let us describe the algorithm.
\begin{enumerate}

\item It takes the context $(U,M,R,I)$ with all ratings, and a target user $u_t$ . It also requires $left\_border$ and $right\_border$ for preferred ratings, i.e. if one wants to get all movies rated in range from 4 to 5, then  $left\_border=4$ and $right\_border=5$. 

\item 	Define the set of movies $M_t=\{m_{t_1}, \ldots, m_{t_q} \}$ that the target user $u_t$ liked, i.e. the ones that she evaluated in the range  $[left\_border,  right\_border]$.

\item	For each movie $m_{t_i} \in M_t$ apply eq.~\ref{op2}. and find the set of users that liked the movie $A_{t_i }=[left\_border, right\_border]_{m_{t_i }}^\square$  for $1\leq i \leq q$. 
As a result one has the set of user subsets: $\{A_{t_1}, \ldots, A_{t_q }\}$.

\item	For each $A_{t_i}  ,1\leq i\leq q$ apply eq.~\ref{op1} to find its description; in our case it is a vector of intervals $d_{t_i}=A_{t_i}^\square=\langle [a_1^{t_i}, b_1^{t_i} ], \ldots, [a_n^{t_i},b_n^{t_i} ] \rangle$    for $1\leq i \leq q$.
Note that,  in case a particular user $u_x$ from  $A_{t_i}$ has not rated $m_y$, i.e. $r_{x,y}= *$, then the algorithm does not take it into account. 

\item	At the last step compute the vector $\mathbf{r}=(R_1,\ldots, R_n ) \in \mathbb{N}^n$ (or $\mathbb{R}^n$ in general case), where 

$$R_j=|\{ i | 1\leq i \leq q, [a_j^{t_i},b_j^{t_i} ]\subseteq [left\_border, right\_border]\}|\mbox{ , i.e. } $$ for each movie $m_j$ the algorithm counts how many of its descriptions $[a_j^{t_i},b_j^{t_i} ]$ are in  $[left\_border,right\_border]$. If $R_j>0$, then the algorithm recommends watching the movie.

\end{enumerate}

Top-$N$ movies with the highest ratings can be selected in similar way.

Let us shortly discuss the time computational complexity. Step 2 requires  $O(|M|)$ operations, steps 3, 4 and 5 perform within $O(|M||U|)$ each. Therefore, the algorithm time complexity is bounded by $O(|M||U|)$.

\textbf{Example 2}

Consider execution of RAPS on the tiny dataset from Table~\ref{tab:RAPSex}.

\begin{table}
\caption{Example of data for RAPS}\label{tab:RAPSex}
	\centering
		\begin{tabular}{|c|c|c|c|c|c|c|}
		\hline
user$\backslash$movie &	$m_1$	 & $m_2$ &	$m_3$ &	 $m_4$	& $m_5$ &	 $m_6$\\
	\hline
$u_1$ &	5	& 3	& 1	& 3 &	5 &	3\\
$u_2$	& 4	& 4 &	1 &	5 &	4 &	3\\
$u_3$	& 5 &	*	 & * &	3 & 	* &	4\\
$u_4$	& * &	3	 & 4	 & *  &	2 &	4\\
$u_5$	& 4 &	* &	4 & 	5 &	4 &	*\\
$u_6$	& 3	& 4	& 5 &	5 &	* &	3\\
$u_7$	& 5 &	4 &	2	 & * &	*	& *\\

			\hline
		\end{tabular}

\end{table}

Let us find a recommendation for user $u_7$.
\begin{enumerate}
\item The input of the algorithm: $t=7$, $left\_border=4$ and $right\_border=5$.
\item  $	M_7=\{m_1,m_2 \}$
\item  $	A_1=[4,5]_{m_1}^\square=\{u_1,u_2,u_3,u_5\} $

$A_2=[4,5]_{m_2}^\square=\{u_2,u_6\} $

\item  

\begin{multline*}
d_1=A_1^\square=\langle [a_1^1,b_1^1 ],[a_2^1,b_2^1 ],[a_3^1,b_3^1 ],[a_4^1,b_4^1 ],[a_5^1,b_5^1 ],[a_6^1,b_6^1 ] \rangle   
=\\
=\langle [4,5],[3,4],[1,4],[3,5],[4,5],[3,4] \rangle 
\end{multline*}

For example, interval $[a_6^1, b_6^1 ]$ is found as follows: 
\begin{multline*}
[a_6^1,b_6^1 ]=[\min(r_{1,6},r_{2,6},r_{3,6},r_{5,6} ), \max(r_{1,6},r_{2,6},r_{3,6},r_{5,6} ) ]=\\
=[\min(3,3,4,*),\max(3,3,4,*) ]=[\min(3,3,4),\max(3,3,4) ]=[3,4].
\end{multline*}
The rest intervals are found in similar way.

$d_2=A_2^\square=\langle [3,4],[4,4],[1,5],[5,5],[4,4],[3,3] \rangle$
   
\item  Taking into account the left and right bounds, the algorithm recommends movies $m_1$ and $m_5$ from $d_1$ and  $m_2$, $m_4$ and $m_5$ from $d_2$.
Therefore  $R=(1,1,0,1,2,0)$, i.e. without already assessed movies by $u_7$, we recommend her to watch $m_4$ and $m_5$.
\end{enumerate}

\section{Experimental evaluation}\label{sec:exp}

\subsection{Data}

For our experimentation we have used freely available data from MovieLens website\footnote{\url{http://grouplens.org/datasets/movielens/}}. The data collection was gathered within The GroupLens Research Project of Minnesota University in 1997--1998. 
The data contains 100000 ratings for 1682 movies by 943 different users. Each user rated no less than 20 movies. That is we have 100000 tuples in the form:
 
user id | item id | rating | timestamp.

Each tuple shows  user id, movie id, the rating she gave to the movie and time when it happened.

\subsection{Quality assessment}

%
%
%

Firstly, for quality assessment of Slope One and RAPS  we used precision and recall measures. Note that we cannot use Mean Absolute Error (MAE) directly, since RAPS actually assume a whole interval like [4,5] for a particular movie, not a number. We select 20\% of users to form our test set and for each test user we split her rated movies into two parts: the visible set and the hidden set.  The first set consists of  80\% rated movies, and the second one contains the remaining 20\%. Moreover, to make the comparison more realistic, movies from the first set were evaluated earlier than those from the second one. It means that first we sort all ratings of a given user by timestamp and then perform splitting.

There is a more general testing scheme based on bimodal cross-validation from \cite{Ignatov:2012q}, which seems to us the most natural and realistic: users from the test set keep only $x\%$ of their rated movies, and the remaining $y\%$ of their ratings are hidden. Thus, by considering each test user in this way, we model a real user whose ratings to other movies are not yet clear, but at the same time we have all ratings' information about the training set of users. In other words, we hide only rectangle of size $x\%$ of test users by $y\%$ of hidden items. One can vary $x$ and $y$ during the investigation of the behaviour of methods under comparison, where the size of top-N recommended list is set to be equal to $y\%$. The part of hidden items can be selected randomly or by timestamp (preferably for realistic scenario). Note that there is no a gold standard approach to test recommender systems, however, there are validated sophisticated schemes \cite{Cremonesi:2010}. The main reason is the following: with only off-line data in hands we cannot verify whether the user will like a not yet seen movie irrespective of assumption that she has seen our recommendation. However, for real systems there is a remedy such as A/B testing, which is applicable only in online setting \cite{Radlinski:2013}.

The adjusted precision and recall are defined below:

\begin{eqnarray}
precision=\frac{|\{relevant \ movies\}\cap\{retrieved \ movies\}\cap \{test \ movies\}|}{|\{retrieved \ movies\}\cap \{test \ movies\}|} \label{prec}	\\
recall=\frac{|\{relevant \ movies\}\cap\{retrieved \ movies\} \cap \{test \ movies\}|}{|\{relevant \ movies\} \cap \{test \ movies\}|} \label{rec}
\end{eqnarray}

These measures allow us to avoid the uncertainty since we do not know how actually a particular user would assess a recommended movie. However, in real recommender system, we would rather ask a user whether the recommendation was relevant, but in our off-line quality assessment scheme we cannot do that. In other words, we assume that for a test user at the moment of assessment there are no movies except the training  and test ones.  

Another issue, which is often omitted in papers on recommender systems, is how to avoid uncertainty when denominators in Precision and Recall are equal to zero (not necessarily simultaneously).

To define the mesaures precisely based on the peculiarities of recommendation task and common sense, we use two types of the definitions for cases when the sets of retrieved and relevant movies for particular user and recommender are empty.

Precision and Recall of the first type are defined as follows:

\begin{itemize}

\item If the sets of relevant movies and retrieved ones are empty, then $Precision=0$ and $Recall=1$.

\item If the set of relevant movies is empty, but the set of retrieved ones is not, then $Precision=0$ and $Recall=1$.
\item If we have the non-empty set of relevant movies, but the set of retrieved movies is empty, then $Precision=0$ and $Recall=0$.

\end{itemize}

Precision of the second type is less tough, but Recall remains the same:

\begin{itemize}

\item If the sets of relevant movies and retrieved ones are empty, then $Precision=1$     
        (since we should not recommend any movie and the recommender has not recommended anything).
				
\item If the set of relevant movies is empty, but the set of retrieved ones is not, then $Precision=0$.

\item If we have the non-empty set of relevant movies, but the set of retrieved movies is empty, then $Precision=1$ (since the recommender has not recommended anything, its output does not contain any non-relevant movie).

\end{itemize}


\subsection{Results}
We have performed three series of tests:
\begin{enumerate}
	\item A movie is worth to watch if its predicted  mark is 5 (i.e. it is [5,5]).
	\item A mark is good if it is from [4,5].
	\item Any mark from [3,5] is good.
	
\end{enumerate}	

All the tests are performed in OS X 10.9.3 with Intel Core i7 2.3 GHz and 8 Gb of memory. The algorithm were implemented in MATLAB -- R2013a.
The results  are presented in Table~\ref{tab:Results}. Note that the reported Precision and Recall are of the first type.

\begin{table}
\caption{RAPS vs Slope One Results}\label{tab:Results}
	\centering
		\begin{tabular}{|c|c|c|c|c|c|}
		\hline
			Algorithm & Preference  & Average &	Average  &	Average &  F1-measure\\
				name & Interval & time, s &	precision &	 recall &\\
			\hline
			
RAPS & [5,5] &	\textbf 3.62	& \textbf 19.42 &	\textbf 50.52 & 28.06\\
Slope One& [5,5] &	18.90  &	1.57 &	23.41& 2.94\\
\hline
RAPS & [4,5] &	\textbf 18.23 &	\textbf 55.61 &	\textbf 63.33 & 59.22\\
Slope One& [4,5] &	18.90	& 53.99	& 30.39 & 38.89 \\
\hline
RAPS & [3,5]	& 32.98	& 80.11	& \textbf 83.65 & 81.84\\
Slope One & [3,5] &	\textbf 18.90 & \textbf	83.81	& 81.88 & 82.83\\

			\hline
		\end{tabular}
	
\end{table}

The criteria are average execution time in seconds, average precision and recall. 
From the table one can see RAPS is drastically better than Slope One by the whole set of criteria in [5,5].
For [4,5] interval both approaches have comparable time and precision, but Slope One has two times lower recall.
For [3,5] interval the algorithms demonstrate similar values of precision and recall but RAPS 1.5 times slower on average.

However, since the compared approaches are different from the output point view (RAPS provides the user with an interval of possible ratings but SlopeOne does it by a single real number), we perform thorough comparison varying the lower bound of acceptable recommendations and using both types of the adjusted precision and recall measures.

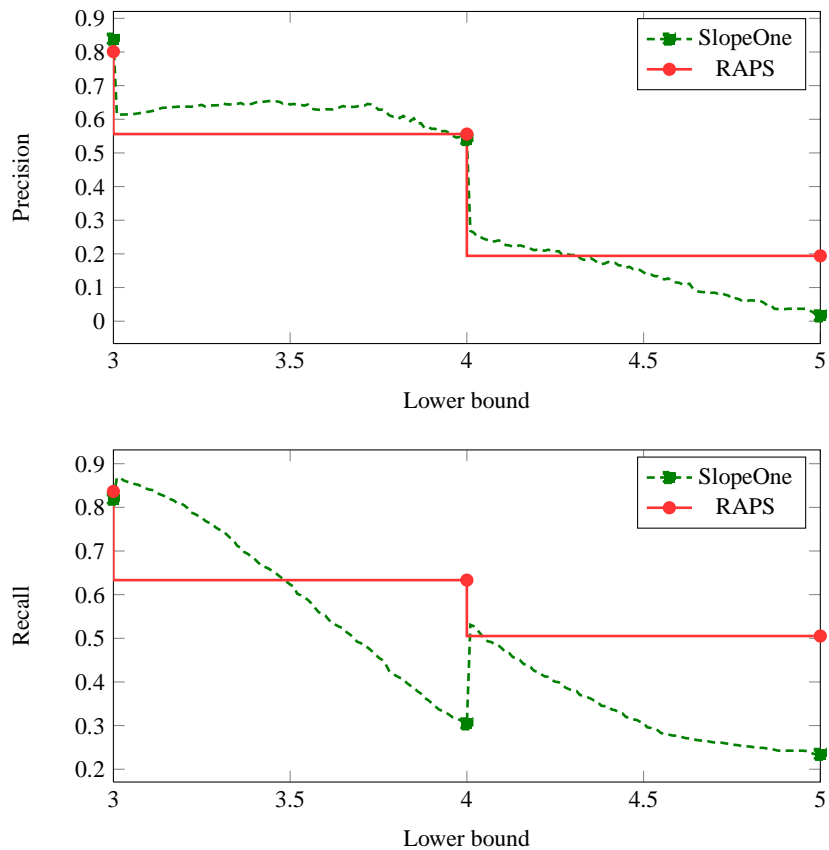
\begin{figure*}[!htb]
\begin{center}
\begin{tikzpicture}
    \begin{axis}[width=0.9\textwidth, height=6cm, xmin=3, xmax=5, xlabel={\footnotesize Lower bound}, ylabel={\footnotesize Precision},
          xticklabels=  {3,3.5,4,4.5,5},
          xtick=        {3,3.5,4,4.5,5},
          yticklabels=  {0,0.1,0.2,0.3,0.4,0.5,0.6,0.7,0.8,0.9,1},
          ytick=        {0,0.1,0.2,0.3,0.4,0.5,0.6,0.7,0.8,0.9,1},
          x tick label style={font=\footnotesize}, y tick label style={font=\footnotesize}
          ]
         \addplot[line width=1.0pt,green!50!black, densely dashed, mark=square*, mark size=2pt, mark repeat=100] table {
3	0.838116836
3.01000000000000	0.612085708
3.02000000000000	0.613818283
3.03000000000000	0.61413088
3.04000000000000	0.614590767
3.05000000000000	0.615469002
3.06000000000000	0.615891768
3.07000000000000	0.617463336
3.08000000000000	0.619484235
3.09000000000000	0.620885502
3.10000000000000	0.622565373
3.11000000000000	0.624449373
3.12000000000000	0.627735449
3.13000000000000	0.631190991
3.14000000000000	0.634343513
3.15000000000000	0.63444721
3.16000000000000	0.634077029
3.17000000000000	0.636385044
3.18000000000000	0.637095083
3.19000000000000	0.637892843
3.20000000000000	0.637394948
3.21000000000000	0.637629626
3.22000000000000	0.637333197
3.23000000000000	0.638526264
3.24000000000000	0.640555123
3.25000000000000	0.642796982
3.26000000000000	0.637700168
3.27000000000000	0.639086031
3.28000000000000	0.641198668
3.29000000000000	0.641608122
3.30000000000000	0.641850534
3.31000000000000	0.646921129
3.32000000000000	0.644403944
3.33000000000000	0.644081955
3.34000000000000	0.645088565
3.35000000000000	0.645698609
3.36000000000000	0.648285634
3.37000000000000	0.643956058
3.38000000000000	0.643905809
3.39000000000000	0.645362939
3.40000000000000	0.649000702
3.41000000000000	0.652585931
3.42000000000000	0.651773015
3.43000000000000	0.652714658
3.44000000000000	0.654107885
3.45000000000000	0.654327682
3.46000000000000	0.653492163
3.47000000000000	0.651686839
3.48000000000000	0.646623404
3.49000000000000	0.648642458
3.50000000000000	0.643871874
3.51000000000000	0.64547084
3.52000000000000	0.643123222
3.53000000000000	0.64286504
3.54000000000000	0.640664292
3.55000000000000	0.643863409
3.56000000000000	0.631994988
3.57000000000000	0.62934148
3.58000000000000	0.628358368
3.59000000000000	0.628097127
3.60000000000000	0.629532811
3.61000000000000	0.629619851
3.62000000000000	0.629113534
3.63000000000000	0.632686364
3.64000000000000	0.635335272
3.65000000000000	0.639352822
3.66000000000000	0.636179314
3.67000000000000	0.636879762
3.68000000000000	0.636749119
3.69000000000000	0.638271243
3.70000000000000	0.641654638
3.71000000000000	0.640755143
3.72000000000000	0.645739973
3.73000000000000	0.643566371
3.74000000000000	0.638334466
3.75000000000000	0.628376397
3.76000000000000	0.628690797
3.77000000000000	0.624371638
3.78000000000000	0.612013107
3.79000000000000	0.60801722
3.80000000000000	0.610242152
3.81000000000000	0.603398489
3.82000000000000	0.610046276
3.83000000000000	0.602632419
3.84000000000000	0.594473172
3.85000000000000	0.602777436
3.86000000000000	0.587729977
3.87000000000000	0.588569557
3.88000000000000	0.574540297
3.89000000000000	0.577644663
3.90000000000000	0.571659254
3.91000000000000	0.571838924
3.92000000000000	0.563799514
3.93000000000000	0.564425743
3.94000000000000	0.564867558
3.95000000000000	0.559433798
3.96000000000000	0.550587318
3.97000000000000	0.545899788
3.98000000000000	0.547128023
3.99000000000000	0.542070931
4	0.539885491
4.01000000000000	0.268644632
4.02000000000000	0.262901268
4.03000000000000	0.249872224
4.04000000000000	0.248955469
4.05000000000000	0.244166787
4.06000000000000	0.24083627
4.07000000000000	0.236028958
4.08000000000000	0.23757686
4.09000000000000	0.240111254
4.10000000000000	0.236082683
4.11000000000000	0.226058197
4.12000000000000	0.225295124
4.13000000000000	0.221394203
4.14000000000000	0.222890539
4.15000000000000	0.225407986
4.16000000000000	0.222563364
4.17000000000000	0.217748857
4.18000000000000	0.21544857
4.19000000000000	0.211156064
4.20000000000000	0.212644678
4.21000000000000	0.21180989
4.22000000000000	0.20936902
4.23000000000000	0.212357321
4.24000000000000	0.209698736
4.25000000000000	0.207233004
4.26000000000000	0.208052724
4.27000000000000	0.196838199
4.28000000000000	0.198365629
4.29000000000000	0.197730858
4.30000000000000	0.195998023
4.31000000000000	0.200177794
4.32000000000000	0.187580343
4.33000000000000	0.182577938
4.34000000000000	0.182094471
4.35000000000000	0.187372878
4.36000000000000	0.181677928
4.37000000000000	0.174148329
4.38000000000000	0.169267231
4.39000000000000	0.171408829
4.40000000000000	0.176170824
4.41000000000000	0.17300042
4.42000000000000	0.172864136
4.43000000000000	0.165706861
4.44000000000000	0.165567237
4.45000000000000	0.159129847
4.46000000000000	0.16176447
4.47000000000000	0.154884174
4.48000000000000	0.155627053
4.49000000000000	0.152018404
4.50000000000000	0.144951022
4.51000000000000	0.13964112
4.52000000000000	0.139203661
4.53000000000000	0.134240194
4.54000000000000	0.133534727
4.55000000000000	0.126087118
4.56000000000000	0.124411633
4.57000000000000	0.127057136
4.58000000000000	0.119882212
4.59000000000000	0.115473041
4.60000000000000	0.114304611
4.61000000000000	0.109473419
4.62000000000000	0.109179797
4.63000000000000	0.111384383
4.64000000000000	0.095996376
4.65000000000000	0.089382619
4.66000000000000	0.087178034
4.67000000000000	0.086296199
4.68000000000000	0.086840629
4.69000000000000	0.084435626
4.70000000000000	0.084729571
4.71000000000000	0.082084068
4.72000000000000	0.084131183
4.73000000000000	0.078840178
4.74000000000000	0.076194675
4.75000000000000	0.066494499
4.76000000000000	0.068184681
4.77000000000000	0.065539179
4.78000000000000	0.059366339
4.79000000000000	0.059366339
4.80000000000000	0.061971758
4.81000000000000	0.061530841
4.82000000000000	0.060869466
4.83000000000000	0.060340365
4.84000000000000	0.05504936
4.85000000000000	0.047112852
4.86000000000000	0.047490781
4.87000000000000	0.03690877
4.88000000000000	0.036067019
4.89000000000000	0.03600823
4.90000000000000	0.03600823
4.91000000000000	0.036978248
4.92000000000000	0.036978248
4.93000000000000	0.036978248
4.94000000000000	0.036978248
4.95000000000000	0.036978248
4.96000000000000	0.036904762
4.97000000000000	0.031613757
4.98000000000000	0.026322751
4.99000000000000	0.015740741
5	0.015740741
				};\label{gr:prec1SlopeOne}
      \addplot[line width=1.0pt,red!80!white,solid,mark=*, mark size=2pt, mark repeat=2] table {
3 0.8011    
3 0.5561
4 0.5561
4 0.1942
5 0.1942
			};\label{gr:prec1RAPS}
     \legend{SlopeOne \\ RAPS \\}
    \end{axis}
\end{tikzpicture}

\

\begin{tikzpicture}
    \begin{axis}[width=0.9\textwidth, height=6cm, xmin=3, xmax=5, xlabel={\footnotesize Lower bound}, ylabel={\footnotesize Recall},
          xticklabels=  {3,3.5,4,4.5,5},
          xtick=        {3,3.5,4,4.5,5},
          yticklabels=  {0,0.1,0.2,0.3,0.4,0.5,0.6,0.7,0.8,0.9,1},
          ytick=        {0,0.1,0.2,0.3,0.4,0.5,0.6,0.7,0.8,0.9,1},
          x tick label style={font=\footnotesize}, y tick label style={font=\footnotesize}
          ]
         \addplot[line width=1.0pt,green!50!black, densely dashed, mark=square*, mark size=2pt, mark repeat=100] table {
3	0.818758409
3.01000000000000	0.868193073
3.02000000000000	0.867003684
3.03000000000000	0.861305804
3.04000000000000	0.858449704
3.05000000000000	0.856449363
3.06000000000000	0.853512846
3.07000000000000	0.851839124
3.08000000000000	0.847430287
3.09000000000000	0.843956304
3.10000000000000	0.840936951
3.11000000000000	0.839722795
3.12000000000000	0.835296234
3.13000000000000	0.830398814
3.14000000000000	0.827386136
3.15000000000000	0.823891762
3.16000000000000	0.819114114
3.17000000000000	0.816143562
3.18000000000000	0.810068094
3.19000000000000	0.809211455
3.20000000000000	0.805004786
3.21000000000000	0.797983455
3.22000000000000	0.788118885
3.23000000000000	0.784968953
3.24000000000000	0.780671583
3.25000000000000	0.776494183
3.26000000000000	0.769111949
3.27000000000000	0.764819822
3.28000000000000	0.761456386
3.29000000000000	0.753734746
3.30000000000000	0.748927757
3.31000000000000	0.745614523
3.32000000000000	0.738113944
3.33000000000000	0.73040763
3.34000000000000	0.724067911
3.35000000000000	0.712358706
3.36000000000000	0.707041508
3.37000000000000	0.697756927
3.38000000000000	0.693527834
3.39000000000000	0.68764993
3.40000000000000	0.681403815
3.41000000000000	0.673133811
3.42000000000000	0.665857455
3.43000000000000	0.663100214
3.44000000000000	0.659419322
3.45000000000000	0.653797583
3.46000000000000	0.648547104
3.47000000000000	0.642154023
3.48000000000000	0.635567796
3.49000000000000	0.629735645
3.50000000000000	0.623229766
3.51000000000000	0.618508153
3.52000000000000	0.602252429
3.53000000000000	0.59868242
3.54000000000000	0.59352831
3.55000000000000	0.588305827
3.56000000000000	0.580201557
3.57000000000000	0.570018644
3.58000000000000	0.560490728
3.59000000000000	0.555100513
3.60000000000000	0.551636899
3.61000000000000	0.541814353
3.62000000000000	0.535647238
3.63000000000000	0.530246168
3.64000000000000	0.526142107
3.65000000000000	0.523138261
3.66000000000000	0.513881253
3.67000000000000	0.508414341
3.68000000000000	0.500209563
3.69000000000000	0.491985829
3.70000000000000	0.489074637
3.71000000000000	0.482247136
3.72000000000000	0.478628245
3.73000000000000	0.471660338
3.74000000000000	0.464101315
3.75000000000000	0.456069573
3.76000000000000	0.450727069
3.77000000000000	0.441242482
3.78000000000000	0.425371087
3.79000000000000	0.419260274
3.80000000000000	0.41393936
3.81000000000000	0.409787042
3.82000000000000	0.406403081
3.83000000000000	0.397628912
3.84000000000000	0.39175889
3.85000000000000	0.387462053
3.86000000000000	0.378888617
3.87000000000000	0.373398288
3.88000000000000	0.363349521
3.89000000000000	0.358821309
3.90000000000000	0.35184222
3.91000000000000	0.344788128
3.92000000000000	0.337392209
3.93000000000000	0.33347764
3.94000000000000	0.330448173
3.95000000000000	0.324872895
3.96000000000000	0.318797254
3.97000000000000	0.315759165
3.98000000000000	0.313207071
3.99000000000000	0.306881131
4	0.303862821
4.01000000000000	0.530823037
4.02000000000000	0.527407196
4.03000000000000	0.518696257
4.04000000000000	0.508706986
4.05000000000000	0.496361307
4.06000000000000	0.493048347
4.07000000000000	0.491156412
4.08000000000000	0.485069466
4.09000000000000	0.483065297
4.10000000000000	0.476272883
4.11000000000000	0.467332764
4.12000000000000	0.465861986
4.13000000000000	0.456997478
4.14000000000000	0.453929681
4.15000000000000	0.450072924
4.16000000000000	0.441454766
4.17000000000000	0.43851758
4.18000000000000	0.434102327
4.19000000000000	0.425785952
4.20000000000000	0.422611349
4.21000000000000	0.41764286
4.22000000000000	0.412836967
4.23000000000000	0.411385949
4.24000000000000	0.404654585
4.25000000000000	0.400415467
4.26000000000000	0.399309096
4.27000000000000	0.39268996
4.28000000000000	0.386653404
4.29000000000000	0.383029644
4.30000000000000	0.380963712
4.31000000000000	0.379521146
4.32000000000000	0.369906942
4.33000000000000	0.368795064
4.34000000000000	0.36511541
4.35000000000000	0.362170618
4.36000000000000	0.357424103
4.37000000000000	0.354602234
4.38000000000000	0.34553581
4.39000000000000	0.344521701
4.40000000000000	0.343542518
4.41000000000000	0.336834497
4.42000000000000	0.335201535
4.43000000000000	0.329537522
4.44000000000000	0.327982926
4.45000000000000	0.319693684
4.46000000000000	0.31774735
4.47000000000000	0.31419288
4.48000000000000	0.313128597
4.49000000000000	0.308901955
4.50000000000000	0.306140056
4.51000000000000	0.296146699
4.52000000000000	0.29464529
4.53000000000000	0.291206137
4.54000000000000	0.289322654
4.55000000000000	0.282497327
4.56000000000000	0.282016326
4.57000000000000	0.279723557
4.58000000000000	0.277478888
4.59000000000000	0.276817513
4.60000000000000	0.275775276
4.61000000000000	0.273276745
4.62000000000000	0.271028993
4.63000000000000	0.270788493
4.64000000000000	0.268583907
4.65000000000000	0.266956522
4.66000000000000	0.26601493
4.67000000000000	0.264692178
4.68000000000000	0.264488678
4.69000000000000	0.262671333
4.70000000000000	0.261348582
4.71000000000000	0.260867581
4.72000000000000	0.260867581
4.73000000000000	0.25954483
4.74000000000000	0.258222079
4.75000000000000	0.255787449
4.76000000000000	0.255787449
4.77000000000000	0.254729248
4.78000000000000	0.252524663
4.79000000000000	0.252284162
4.80000000000000	0.252080662
4.81000000000000	0.250669727
4.82000000000000	0.249547393
4.83000000000000	0.249106476
4.84000000000000	0.248876432
4.85000000000000	0.247465497
4.86000000000000	0.247465497
4.87000000000000	0.244643628
4.88000000000000	0.242676459
4.89000000000000	0.242472959
4.90000000000000	0.242472959
4.91000000000000	0.242472959
4.92000000000000	0.242472959
4.93000000000000	0.242472959
4.94000000000000	0.242472959
4.95000000000000	0.242472959
4.96000000000000	0.241828542
4.97000000000000	0.240505791
4.98000000000000	0.235214785
4.99000000000000	0.234092451
5	0.234092451
				};\label{gr:rec1SlopeOne}
      \addplot[line width=1.0pt,red!80!white,solid,mark=*, mark size=2pt, mark repeat=2] table {
3 0.8365
3 0.6333
4 0.6333
4 0.5052
5 0.5052
			};\label{gr:rec1RAPS}
    \legend{SlopeOne \\ RAPS \\}
    \end{axis}
\end{tikzpicture}

\caption{Precision and Recall of the first type for RAPS and SlopeOne for the varying lower bound}\label{prec-rec1}

\end{center}
\end{figure*}

From Fig.~\ref{prec-rec1} one can conclude that RAPS dominates SlopeOne in most cases by Recall. As for Precision, even though for [5,5] interval RAPS is significantly better, after lower bound of 4.4 SlopeOne shows comparable but slightly better Precision in most cases.

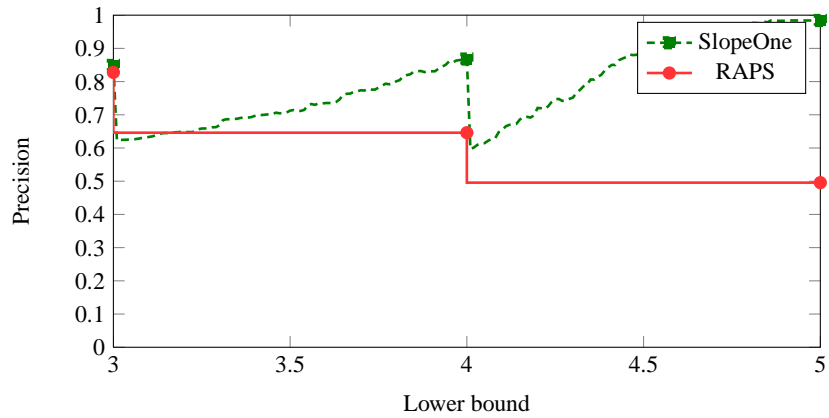
\begin{figure*}[!htb]
\begin{center}


\begin{tikzpicture}
    \begin{axis}[width=0.9\textwidth, height=6cm, xmin=3, xmax=5, xlabel={\footnotesize Lower bound}, ymin=0, ymax=1, ylabel={\footnotesize Precision},
          xticklabels=  {3,3.5,4,4.5,5},
          xtick=        {3,3.5,4,4.5,5},
          yticklabels=  {0,0.1,0.2,0.3,0.4,0.5,0.6,0.7,0.8,0.9,1},
          ytick=        {0,0.1,0.2,0.3,0.4,0.5,0.6,0.7,0.8,0.9,1},
          x tick label style={font=\footnotesize}, y tick label style={font=\footnotesize}
          ]
         \addplot[line width=1.0pt,green!50!black, densely dashed, mark=square*, mark size=2pt, mark repeat=100] table {
				3	0.848698846270952
3.01000000000000	0.622667718488185
3.02000000000000	0.624400293972512
3.03000000000000	0.624712890247240
3.04000000000000	0.625172777156739
3.05000000000000	0.626051012901208
3.06000000000000	0.626473778126484
3.07000000000000	0.628045346922130
3.08000000000000	0.630066245203660
3.09000000000000	0.631467512993947
3.10000000000000	0.633147383190538
3.11000000000000	0.635031383687813
3.12000000000000	0.638317459987144
3.13000000000000	0.641773001116216
3.14000000000000	0.644925523843338
3.15000000000000	0.645029220987046
3.16000000000000	0.644659039475552
3.17000000000000	0.646967054972160
3.18000000000000	0.647677093368142
3.19000000000000	0.648474853671873
3.20000000000000	0.647976958843581
3.21000000000000	0.648211636323990
3.22000000000000	0.647915207329567
3.23000000000000	0.649108274681489
3.24000000000000	0.656428138941236
3.25000000000000	0.658669998158593
3.26000000000000	0.658864188706741
3.27000000000000	0.660250051986074
3.28000000000000	0.662362688835218
3.29000000000000	0.662772143274170
3.30000000000000	0.668305560409499
3.31000000000000	0.683958165771909
3.32000000000000	0.686731986628871
3.33000000000000	0.686409997491978
3.34000000000000	0.687416607461619
3.35000000000000	0.688026650889519
3.36000000000000	0.690613676741227
3.37000000000000	0.691575105192074
3.38000000000000	0.691524856783466
3.39000000000000	0.692981986813228
3.40000000000000	0.696619749969744
3.41000000000000	0.700204978399153
3.42000000000000	0.699392062923348
3.43000000000000	0.700333705226241
3.44000000000000	0.701726932432082
3.45000000000000	0.701946729539072
3.46000000000000	0.706402215686345
3.47000000000000	0.704596891553757
3.48000000000000	0.704824462130890
3.49000000000000	0.706843516521075
3.50000000000000	0.712654942450839
3.51000000000000	0.714253909272609
3.52000000000000	0.711906290660443
3.53000000000000	0.711648109050170
3.54000000000000	0.714738366553644
3.55000000000000	0.728519493364304
3.56000000000000	0.732524088635674
3.57000000000000	0.729870580933032
3.58000000000000	0.734178473991448
3.59000000000000	0.733917232892273
3.60000000000000	0.735352916509632
3.61000000000000	0.735439957156685
3.62000000000000	0.734933639568447
3.63000000000000	0.738506469646042
3.64000000000000	0.746446382758931
3.65000000000000	0.761045944007480
3.66000000000000	0.763163441164851
3.67000000000000	0.763863888536380
3.68000000000000	0.769024251593650
3.69000000000000	0.770546374775446
3.70000000000000	0.773929770104748
3.71000000000000	0.773030275722477
3.72000000000000	0.778015105399690
3.73000000000000	0.775841503591684
3.74000000000000	0.775900603780875
3.75000000000000	0.781815550721518
3.76000000000000	0.787420955903456
3.77000000000000	0.793683807765856
3.78000000000000	0.791907287178206
3.79000000000000	0.793202405671762
3.80000000000000	0.800718342649369
3.81000000000000	0.804456689840097
3.82000000000000	0.816395482451158
3.83000000000000	0.819563635902174
3.84000000000000	0.821986399915779
3.85000000000000	0.830290663297098
3.86000000000000	0.831116220794476
3.87000000000000	0.831955800240990
3.88000000000000	0.828508551429836
3.89000000000000	0.831612917177382
3.90000000000000	0.830918513437325
3.91000000000000	0.831098183700464
3.92000000000000	0.838931788925915
3.93000000000000	0.844849023161002
3.94000000000000	0.850581843817471
3.95000000000000	0.861021099148848
3.96000000000000	0.857465625189059
3.97000000000000	0.863360104963881
3.98000000000000	0.864588340396330
3.99000000000000	0.870113258835600
4	0.867927819003102
4.01000000000000	0.596686960398621
4.02000000000000	0.601525606272765
4.03000000000000	0.609660584054820
4.04000000000000	0.614034834026341
4.05000000000000	0.614537157406511
4.06000000000000	0.621788651257282
4.07000000000000	0.627563349973403
4.08000000000000	0.629111251883624
4.09000000000000	0.647518661517663
4.10000000000000	0.654072101311680
4.11000000000000	0.665211636393424
4.12000000000000	0.669739568896687
4.13000000000000	0.671129652945501
4.14000000000000	0.672625988380851
4.15000000000000	0.685725446433725
4.16000000000000	0.693462835193948
4.17000000000000	0.693939333271662
4.18000000000000	0.691639046441209
4.19000000000000	0.697928550970879
4.20000000000000	0.720581186274013
4.21000000000000	0.719746398417750
4.22000000000000	0.717305527509755
4.23000000000000	0.725584834323593
4.24000000000000	0.738799265395167
4.25000000000000	0.746915543390800
4.26000000000000	0.747735263491566
4.27000000000000	0.741811743804471
4.28000000000000	0.743339173821630
4.29000000000000	0.747995408239769
4.30000000000000	0.751553578464606
4.31000000000000	0.766315359893054
4.32000000000000	0.774881930239073
4.33000000000000	0.785752541109684
4.34000000000000	0.790560079250555
4.35000000000000	0.806420497491926
4.36000000000000	0.806016552727266
4.37000000000000	0.814359969145217
4.38000000000000	0.820060881988987
4.39000000000000	0.832784489950690
4.40000000000000	0.842837490258125
4.41000000000000	0.850249096876081
4.42000000000000	0.850112813406464
4.43000000000000	0.864119559159242
4.44000000000000	0.869270940699512
4.45000000000000	0.873415561510800
4.46000000000000	0.881341189674523
4.47000000000000	0.879751898987488
4.48000000000000	0.880494777508144
4.49000000000000	0.887468139243411
4.50000000000000	0.885691762675890
4.51000000000000	0.896254876810432
4.52000000000000	0.901108423167247
4.53000000000000	0.901435961590023
4.54000000000000	0.906021499508894
4.55000000000000	0.909155901417806
4.56000000000000	0.918062426990998
4.57000000000000	0.920707929636501
4.58000000000000	0.924115016376921
4.59000000000000	0.924996850592089
4.60000000000000	0.923828420256992
4.61000000000000	0.934870244394054
4.62000000000000	0.934576622671861
4.63000000000000	0.942072213500785
4.64000000000000	0.942557222319127
4.65000000000000	0.941234470996376
4.66000000000000	0.939029885458457
4.67000000000000	0.943439056534295
4.68000000000000	0.943983485650152
4.69000000000000	0.952160493827160
4.70000000000000	0.952454438565550
4.71000000000000	0.949808935920047
4.72000000000000	0.957147056353406
4.73000000000000	0.957147056353406
4.74000000000000	0.954501553707903
4.75000000000000	0.955383387923070
4.76000000000000	0.957073570168808
4.77000000000000	0.959719072814311
4.78000000000000	0.958837238599143
4.79000000000000	0.964128243890149
4.80000000000000	0.972024668453240
4.81000000000000	0.976874756636662
4.82000000000000	0.976213380975286
4.83000000000000	0.975684280446185
4.84000000000000	0.980975285737191
4.85000000000000	0.978329783091688
4.86000000000000	0.983998717332051
4.87000000000000	0.983998717332051
4.88000000000000	0.983156966490300
4.89000000000000	0.983098177542622
4.90000000000000	0.983098177542622
4.91000000000000	0.984068195179306
4.92000000000000	0.984068195179306
4.93000000000000	0.984068195179306
4.94000000000000	0.984068195179306
4.95000000000000	0.984068195179306
4.96000000000000	0.983994708994709
4.97000000000000	0.983994708994709
4.98000000000000	0.983994708994709
4.99000000000000	0.983994708994709
5 0.983994708994709				
				};\label{gr:prec2SlopeOne}
      \addplot[line width=1.0pt,red!80!white,solid,mark=*, mark size=2pt, mark repeat=2] table {
3 0.8276    
3 0.6461
4 0.6461
4 0.4957
5 0.4957
			};\label{gr:prec2RAPS}
     \legend{SlopeOne \\ RAPS \\}
    \end{axis}
\end{tikzpicture}

\

\caption{Precision of  the second type for RAPS and SlopeOne for the varying lower bound}\label{prec-rec2}

\end{center}
\end{figure*}

From Fig.~\ref{prec-rec2} one can see that SlopeOne is significantly better in terms of precision. Only on the interval [3,5] the difference between SlopeOne and RAPS is negligible (the lower bound value equals 3). The reasonable explanations is as follows:
for SlopeOne there are more cases when $\{retrieved \ movies\}=\emptyset$ irrespective of the size of $\{relevant \ movies\}$. Remember that in such cases Precision of the second type is equal to 1. In other words SlopeOne is really more precise (or even concise): in such cases it just does not recommend anything. However, it can be hardly judged in movie recommendation domain that a recommender is good when it does not recommend.

We can conclude that the proposed recommender technique based on pattern structures has its right to be used. Since the Slope One algorithm was exploited in real recommender systems \cite{Lemire:2005}, we can suggest our technique for usage as well.

\section{Conclusion and further work}\label{sec:end}

In this paper we proposed the technique for movie recommendation based on Pattern Structures (RAPS). Even though this algorithm is oriented to movie recommendations, it can be easily used in other recommender domains where users evaluate items. 

The performed experiments (RAPS vs Slope One) showed that recommender system based on Pattern Structures demonstrates acceptable precision, better recall in most cases and reasonable execution time.

Of course, in future RAPS should be compared with other recommender techniques to make a final conclusion about its applicability. An interplay between interval-based recommendations and regression-like ones deserves a more detailed treatment as well.

The further work can be continued in the following directions:
\begin{enumerate}
\item Further modification and adjustment of RAPS.
\item Development of the second variant of  Pattern Structures based recommender. There is a conjecture that for the second derivation operation (operator Galois from eq.\ref{op2}) being applied to more than one movie with high marks we may obtain relevant predictions as well.
\item Comparison with existing popular techniques, e.g. SVD and SVD++. 
\end{enumerate}

\subsubsection*{Acknowledgments}
We would like to thank Mehdi Kaytoue, Sergei Kuznetsov and Sergei Nikolenko for their comments, remarks and explicit and implicit help during the paper preparations. 
The first author has been supported by the Russian Foundation for Basic Research grants no. 13-07-00504 and 	
14-01-93960 and made a contribution within the project ``Data mining based on applied ontologies and lattices of closed descriptions'' supported by the Basic Research Program of the National Research University Higher School of Economics.
 We also deeply thank the reviewers for their comments and remarks that helped.

\bibliographystyle{splncs}

\bibliography{patrecbib}

\end{document}